\documentclass[12pt]{article}
\begin{document}

\renewcommand{\theequation}{\arabic{section}.\arabic{equation}}
\baselineskip=20pt
\newfont{\largebx}{cmssbx10 scaled\magstep4}   

\title{\largebx States interpolating between number and coherent states
             and their interaction\\ with atomic systems }
\author{{\bf Hongchen Fu\footnote{On leave of absence from Institute
                            of Theoretical Physics, Northeast
                            Normal University, Changchun 130024,
                            P.\,R.\,China. Email:h.fu@open.ac.uk},
         Yinqi Feng\footnote{Email: y.feng@open.ac.uk}  and
         Allan I. Solomon\footnote{Email: a.i.solomon@open.ac.uk}}\\
{\normalsize \it Quantum Processes Group,
                 The Open University,}\\
{\normalsize \it Milton Keynes, MK7 6AA, UK}}

\date{\ }

\maketitle

\begin{abstract}
Using the eigenvalue definition of  binomial states we
construct  new intermediate number-coherent states which
reduce to number and coherent states in two different
limits. We reveal the connection of these intermediate states with
photon-added coherent states and investigate their
non-classical properties and quasi-probability
distributions in detail. It is of interest to note that
these new states, which interpolate between coherent
states and number states, neither of which exhibit
squeezing, are nevertheless squeezed states. A
scheme to produce these states is proposed.
We also study the interaction of these states with atomic systems
in the framework of the two-photon Jaynes-Cummings model,
and describe the response of the atomic system as it varies  between
the pure Rabi oscillation and the collapse-revival mode and
investigate field observables such as photon number
distribution, entropy and the Q-function.
\end{abstract}

\vspace{2cm}
\begin{center} \sf
    J.Phys.A:Math.Gen.33\,(2000)\,in\,press.
\end{center}    

\newpage


\section{Introduction}

Since Stoler, Saleh and Teich proposed the binomial states (BS) in
1985 \cite{Sto}, so-called {\it intermediate} states which interpolate between
some fundamental states such as number states, coherent
and squeezed states and phase states have attracted much
attention \cite{ins}. The BS are finite linear combinations
of number states
\begin{equation}
   |\eta, M\rangle = \sum_{n=0}^M \left[ {M\choose n}\eta^n
   (1-\eta)^{M-n} \right]^{1/2} |n\rangle,
\end{equation}
where $ M$ is a non-negative integer, $ \eta $ is
a {\em real} probability ($ 0<\eta <1$) and
$|n\rangle $ is a number state of the
radiation field. The photon number distribution is clearly
a binomial distribution, whence the name
{\it binomial state}. The BS are intermediate
number-coherent states in the sense that they reduce
to number and coherent states in different limits
\begin{equation}
   |\eta,M\rangle \longrightarrow \left\{
   \begin{array}{ll}
     |M\rangle, &  \eta\rightarrow 1,\\
     |0\rangle,  & \eta\rightarrow 0,\\
     |\alpha\rangle,  & \eta\to 0,\ M\to\infty, \eta M=\alpha^2.
   \end{array}\right.
\end{equation}
The BS also admit an eigenvalue definition \cite{GBS}
\begin{equation}
   \left(\sqrt{\eta}N+\sqrt{1-\eta}\sqrt{M-N}a\right)
   |\eta,M\rangle = \sqrt{\eta} M |\eta, M\rangle,
\end{equation}
where $ a $, $a^\dagger $ and $ N $ are the annihilation,
creation and the number operators, respectively.
The algebra involved is the su(2) algebra
(Holstein-Primakoff realization \cite{HP})
\begin{equation}
   J^+=\sqrt{M-N}a, \ \ \ \ \
   J^-=a^\dagger \sqrt{M-N},\ \ \ \ \
   J^3=\frac{M}{2}-N,
\end{equation}
and in the present case
the coherent state limit is essentially the contraction
of su(2) to the Heisenberg-Weyl algebra generated by
$a^\dagger, a$ and 1.

Since number and coherent states are
eigenstates of the number operator $N$ and the annihilation
operator $ a $, respectively, it would seem more natural that,  to define states interpolating
between number and coherent states, we consider the eigenvalue equation of a linear
combination of $N$ and $ a$ itself (not $J^+$), namely,
\begin{equation}
   \left(\sqrt{\eta}N+\sqrt{1-\eta}a\right)
   \|\eta,\beta\rangle = \beta\|\eta,\beta\rangle.
   \label{equation1}
\end{equation}
Here $ 0<\eta<1 $ as before and $\beta $ is an eigenvalue
which will be determined  not only from the eigenvalue equation
Eq.(\ref{equation1}) but also by a {\it physical requirement}
(see Sec.\,2).

In this paper we study the states
$\|\eta,\beta\rangle $ and their various properties.
We find that for $\beta=
\sqrt{\eta}M$ ($M$ a non-negative integer), the
solutions to Eq.\,(\ref{equation1}) are indeed intermediate
states which interpolate between number and coherent states. We also
find that these states are closely related to the photon-added
coherent states proposed by Agarwal and Tara \cite{Aga}.
The properties of this new state, such as their
sub-Poissonian statistics,
antibunching effects and squeezing effects, as well as their
quasi-probability distributions (the Q and Wigner functions),
are studied in detail. Although coherent and number
states are not squeezed, the new interpolating states
are squeezed, and
exhibit highly nonclassical behavior. We also propose a
scheme to produce these intermediate states in a cavity.

The intermediate number-coherent states are
of particular interest in their interaction with atomic systems.  In the context of the Jaynes-Cummings (JC) model,
atomic population inversion exhibits two completely
different phenomena:
Rabi oscillation  when the
field is initially prepared in a number state; and periodic collapse and revival when the
field is initially prepared in a coherent
state.  We naturally expect that the states proposed
in this paper will present  phenomena intermediate between
Rabi oscillation and periodic collapse-revival,
given that the initial state of the field is in an intermediate
state. In Sec.\,6 we study the interaction of the states
with the atomic systems based on the two-photon JC model
and we indeed observe this intermediate behavior. We also
give an analytic derivation of the approximate photon number distribution
of the field and find it exhibits strong oscillation at
$\tau=\pi/4, 3\pi/4$ ($\tau$ the scaled time). This
phenomena is explained physically in terms of the entropy
and Q-function of the field.

\section{New intermediate number-coherent states}
\setcounter{equation}{0}

In this section we solve the eigenvalue equation
Eq.(\ref{equation1}), discuss the relation of the
states (1.5) to photon-added coherent states and study the
limit to number and coherent states.

\subsection{Solutions}

Expanding the state $\|\eta,\beta\rangle $ in
number states
\begin{equation}
   \|\eta,\beta\rangle = \sum_{n=0}^\infty C_n |n\rangle
   \label{expand}
\end{equation}
inserting it into Eq.\,(\ref{equation1})
and comparing the two sides of the equation, we find the solution of the eigenvalue equation
Eq.(\ref{equation1})
\begin{equation}
   \|\eta, \beta\rangle=C_0\sum_{n=0}^\infty
   \frac{[\beta-\sqrt{\eta}(n-1)][\beta-\sqrt{\eta}(n-2)]
   \cdots \beta}{(\sqrt{1-\eta})^n\sqrt{n!}} |n\rangle,
   \label{solution2}
\end{equation}
where $ C_0$ is the normalization constant.
Here, the eigenvalue $\beta$
is an arbitrary complex number.

It is easy to see that for any complex number $\beta$
the state Eq.(\ref{solution2}) reduces to the coherent state
$|\beta\rangle \equiv e^{-\frac{|\beta|^2}{2}}\sum_{n=0}^\infty
\frac{\beta^n}{\sqrt{n!}} |n\rangle $
in the limit $\eta\to 0 $, as expected.
However, it does not have a number state limit for arbitrary
$\beta$ since number states are eigenstates of $N$ with
non-negative integer eigenvalues. Further, we
would like to have truncated states which are {\em finite}
superpositions of the number states just as the binomial
states are. With this in mind, we must choose
$\beta=\sqrt{\eta}M$, where $M$ is a non-negative integer.
In this case it is easy to see that the coefficients $ C_n $
are truncated
\begin{equation}
   C_n = \left\{
   \begin{array}{ll}
     0, & \mbox{when } n>M, \\
     \left(\sqrt{\displaystyle \frac{\eta}{1-\eta}}\right)^n
     {\displaystyle\frac{M!}{(M-n)!\sqrt{n!}}}
     C_0, &\mbox{when } n\leq M.
   \end{array}\right. \label{trunc1}
\end{equation}
Here the normalization constant $ C_0(\eta,M) $
is obtained as
\begin{equation}
   C_0(\eta,M)=\left[\sum_{n=0}^M\left(\frac{\eta}{1-\eta}
      \right)^n \frac{(M!)^2}{[(M-n)!]^2 n!}\right]^{-
      \frac{1}{2}}=
      \frac{\lambda^M}{\sqrt{M!L_M(-\lambda^2)}},
      \label{trunc2}
\end{equation}
where $\lambda\equiv \sqrt{(1-\eta)/\eta}$ and $L_M(x)$
is the Laguerre polynomial \cite{asso}
\begin{equation}
   L_M(x)=\sum_{n=0}^M\frac{1}{n!}{M\choose M-n}(-1)^n
   x^n.
\end{equation}
Inserting Eq.\,(\ref{trunc1}) and Eq.\,(\ref{trunc2}) into
Eq.\,(\ref{expand}), we obtain the desired solution
$\|\eta,(\beta=\sqrt{\eta}M)\rangle\equiv
\|\eta,M\rangle $
\begin{equation}
   \|\eta,M\rangle =\frac{1}{\sqrt{M!L_M(-\lambda^2)}}
   \sum_{n=0}^M \lambda^{M-n}
   \frac{M!}{(M-n)!\sqrt{n!}}|n\rangle,
   \label{state1}
\end{equation}
which is a {\em finite} linear superposition
of number states.


We now consider the limiting cases of above state Eq.(\ref{state1})
as number and coherent states.
First consider the limit $ \eta\to 1 $. From the
number-state expansion Eq.(\ref{state1}), it follows
that
\begin{equation}
    C_n=\frac{\lambda^{M-n}M!}{\sqrt{M!n!}(M-n)!}
    \longrightarrow \delta_{M,n},
\end{equation}
namely, $\|\eta,M\rangle\to |M\rangle $. Then,
in the different limit $\eta\to 0$, $M\to\infty$
with $\sqrt{\eta}M=\alpha $ a real constant,
we have
\begin{equation}
    \frac{M!}{(M-n)!}\to M^n,\ \ \ \ \
    \lambda^{-n}M^n\to \alpha^n,\ \ \ \ \
    C_0\to \exp(-\alpha^2/2).
\end{equation}
and therefore Eq.(\ref{state1}) reduces to the coherent
state $|\alpha\rangle$.

The above discussion shows that the state $\|\eta,M\rangle $
may be considered as an intermediate state which interpolates
between a number state and a coherent state.

\subsection{Connection with photon-added coherent states}

The state Eq.(\ref{state1}) can be written in more elegant form.
By making use of $|n\rangle = \frac{a^{\dagger n}}{\sqrt{n!}}|0
\rangle $, we can write Eq.(\ref{state1}) as
\begin{equation}
   \|\eta,M\rangle = \frac{1}{\sqrt{M!L_M(-\lambda^2)}}
   \left[\sum_{n=0}^M
   {M\choose n} (a^\dagger )^n \lambda^{M-n}\right]
   |0\rangle = \frac{1}{\sqrt{M!L_M(-\lambda^2)}}
   \left(a^\dagger +\lambda\right)^M
   |0\rangle, \label{BNNN}
\end{equation}
where we have used the binomial formula.

Furthermore, thanks to the equation
(real $\lambda $ in our case)
\begin{equation}
   D(-\lambda)a^\dagger D(\lambda)=a^\dagger +\lambda,
\end{equation}
where $ D(\lambda)$ is the displacement operator
\begin{equation}
   D(\lambda)=\exp\left[\lambda (a^\dagger -a)\right],
\end{equation}
we can rewrite Eq.\,(\ref{BNNN}) in the  form
\begin{eqnarray}
   \|\eta,M\rangle &=& \frac{1}{\sqrt{M!L_M(-\lambda^2)}}
   D(-\lambda)a^{\dagger M}D(\lambda)
   |0\rangle \nonumber \\ &=&
   \frac{1}{\sqrt{M!L_M(-\lambda^2)}}
   D(-\lambda) a^{\dagger M} |\lambda\rangle \nonumber \\
   &\equiv & D(-\lambda)|\lambda,M\rangle
   \label{212}
\end{eqnarray}
where $ |\lambda\rangle=D(\lambda)|0\rangle$ is a
coherent state and
\begin{equation}
   |\lambda,M\rangle \equiv \frac{1}{\sqrt{M!L_M(-\lambda^2)}}
   a^{\dagger M}|\lambda\rangle
   \label{final}
\end{equation}
is a so-called {\em photon-added coherent state}
or  {\em excited coherent state }\cite{Aga}.
So from Eq.\,(\ref{final}) we conclude that our  new intermediate number-coherent states
are {\it displaced excited coherent states}.

However, we would like to point out that our states are very
different from the photon-added coherent states. The
photon-added states are an {\it infinite} superposition of number
states from $ M $ to infinity, while our states are a
{\it finite} superposition of number states from 0 to $ M $.

\section{Nonclassical Properties}
\setcounter{equation}{0}

In  this section we shall investigate the statistical
and squeezing properties of $\|\eta,M\rangle $ , with  special
emphasis on the comparison with those of the BS.

\subsection{Photon statistics}

The easily-derived relation
\begin{equation}
    a^k \|\eta,M\rangle =\left[\frac{M(M-1)\cdots (M-k+1)
    L_{M-k}(-\lambda^2)}{
    L_M(-\lambda^2)}\right]^{1/2} \|\eta,M-k\rangle,
\end{equation}
for $k\leq M$ and $ a^k \|\eta,M\rangle=0 $ for $ k>M $,
gives the mean value of $\langle
N\rangle $ and $ \langle N^2 \rangle $
\begin{eqnarray}
    &&\langle N\rangle =
    \frac{ML_{M-1}(-\lambda^2)}{L_M(-\lambda^2)}, \\
    &&\langle N^2\rangle =
    \frac{M(M-1)L_{M-2}(-\lambda^2)+ML_{M-1}(-\lambda^2)}{
    L_M(-\lambda^2)},
\end{eqnarray}
and Mandel Q-parameter \cite{Qpar}
\begin{equation}
    Q(\eta,M)=\frac{\langle N^2\rangle -\langle N\rangle^2
              }{\langle N\rangle}-1=
    (M-1)\frac{L_{M-2}(-\lambda^2)}{L_{M-1}(-\lambda^2)}-
    M\frac{L_{M-1}(-\lambda^2)}{L_{M}(-\lambda^2)}.
\end{equation}
If $Q(\eta,M)<0$ ({\em resp.} $Q(\eta,M)>0$), the field in the
state $\|\eta,M\rangle$ is sub-Poissonian ({\em resp.}super-Poissonian).
$Q(\eta,M)=0$ corresponds to Poissonian statistics.

For a fixed $ M $, there are two extreme cases, $\eta=0$
(or $\lambda=\infty$) and $\eta=1$ (or $\lambda=0 $).
It is easy to see that
\begin{equation}
    Q(\eta,M)\longrightarrow\left\{
    \begin{array}{ll}
        -1 & \lambda=0,\\
        0  & \lambda\to \infty,
    \end{array}\right.
\end{equation}
in agreement  with the Q-parameter values for number states
and the vacuum state. Here we have
used the fact $L_M(0)=1$ and $ L_m(x)/L_n(x)\to 0 $ for $m<n$
and $x\to\infty$.

Fig. 1 is a plot of $ Q(\eta,M)$ with respect to $\eta $
for $M=2, 50, 100$. The Q-parameter of the binomial states (BS)
is also presented in the figure ($Q=-\eta$ for any $M$). From the
figure we observe that the field in $\|\eta,M\rangle$
is {\em sub-Poissonian} except for the case $\eta=0$.

We say that a field is antibunched if the second-order correlation
function
$g^{(2)}(0)=\langle a^\dagger a^\dagger a a\rangle/
\langle a^\dagger a \rangle^2<1$\cite{anti}. In fact, the occurrence
of antibunching effects and sub-Poissonian statistics
coincides for single mode, time-independent fields such as
the state $\|\eta,M\rangle $ of this paper.
So the field $\|\eta,M\rangle $ is antibunched except
at the point $\eta=0$.

\subsection{Squeezing properties}

Define two quadratures $x$ (coordinate) and $p$ (momentum)
\begin{equation}
    x=\frac{1}{\sqrt{2}}\left(a+a^\dagger\right),\ \ \ \
    p=\frac{1}{\sqrt{2}i}\left(a-a^\dagger\right).
\end{equation}
Then we can easily obtain their variances
$ (\Delta x)^2\equiv \langle x^2\rangle-
\langle x \rangle^2 $ and $ (\Delta p)^2\equiv \langle
p^2\rangle-\langle p \rangle^2 $
\begin{eqnarray}
    ( \Delta x)^2  &=& \frac{1}{2}+
    \frac{ML_{M-1}(-\lambda^2)}{L_M(-\lambda^2)}+
    \frac{\lambda^2 L_{M-2}^{(2)}(-\lambda^2)}{L_M(-\lambda^2)}
    -2\left[\frac{\lambda L_{M-1}^{(1)}(-\lambda^2)}{L_M(-\lambda^2)}
    \right]^2,\\
    (\Delta p)^2 &=& \frac{1}{2}+
    \frac{ML_{M-1}(-\lambda^2)}{L_M(-\lambda^2)}
    -\frac{\lambda^2 L_{M-2}^{(2)}(-\lambda^2)}{L_M(-\lambda^2)},
\end{eqnarray}
where $ L_m^{(k)}(x)$ is the associated Laguerre polynomial defined
by \cite{asso}
\begin{equation}
    L_m^{(k)}(x)=\sum_{n=0}^m \frac{(m+k)!}{(m-n)!n!(k+n)!}
    (-x)^n, \ \ \ \ \ (k>-1).
\end{equation}
If $(\Delta x)^2  <1/2$ (or $(\Delta p)^2<1/2$),
we say the state is {\em squeezed} in
the quadrature $ x $ (or $p$).

Fig.2 is a plot showing how the variance
$(\Delta x)^2$ depends on
the parameters $\eta $ and $ M $. When $\eta=0 $,
$(\Delta x)^2=1/2 $ since the state is just the vacuum
state and in this case the field is not squeezed. Then,
as $\eta $ increases the field becomes squeezed until
maximum squeezing is reached; then the squeezing decreases
until it disappears at a point $\eta_0$
depending on $ M $. We note
that $\eta_0<1$ when $M>0$ since
$(\Delta x^2)=M+1/2>1/2$ when $\eta\to 1$.

We also observe from Fig.2 that the larger $M$, the stronger
the squeezing, and the wider the squeezing range.


It is known that the optimal signal-to-quantum noise
ratio for an arbitrary quantum state
\begin{equation}
    \rho=\frac{\langle x\rangle^2}{(\Delta x)^2}
\end{equation}
has the value $4N_s (N_s+1)$ which is attainable for
the usual coherent
squeezed state\cite{ratio}.  For a coherent
state  the maximal ratio is $4N_s$, where $N_s$ is the mean value of the
number operator $N$ for the quantum state.

For the intermediate number-coherent state
$\|\eta,M\rangle $, the signal-to-quantum noise
ratios for different parameters $\eta $ and $M$
are shown in Fig.\,3. The ratio for
$\eta=0$ and $\eta=1$, which correspond to the
vacuum state and number state respectively,
is zero. For other $\eta$, we find from
Fig.\,3 (a) that the larger $M$,
the larger the ratio. Fig.3 (b) gives plots of
$ 4\langle N\rangle (\langle N\rangle +1)$
($\langle N\rangle $ is given by Eq.(3.2)),
$ 4\langle N\rangle$ and the ratio for the state
$\|\eta,M\rangle $ with $M=10$. We find that

(1) the ratio for $\|\eta,M\rangle $ is always
smaller than the value
$ 4\langle N\rangle (\langle N\rangle +1)$,
which is in accord with the general result\cite{ratio};

(2) for some values of $\eta $ the ratio is
larger than $4\langle N\rangle $.
We observe that the states with ratio larger
than $ 4\langle N\rangle $ correspond to
squeezed states (see Fig.\,2.).

\section{Quasi-probability distributions}
\setcounter{equation}{0}

Quasi-probability distributions \cite{qp} in the coherent
state basis turn out to be useful measures for studying the
nonclassical features of radiation fields. In this section
we study the Q-function (also called the Husimi function)
and the Wigner function of the state
$\|\eta,M\rangle $.

One can prove that (see Appendix A),
if two states $|\psi\rangle_\alpha $ and $|\psi\rangle $
satisfy $|\psi\rangle_\alpha =D(\alpha)|\psi\rangle $,
where $D(\alpha)=e^{\alpha a^\dagger-\alpha^* a}$ is the
displacement operator, the Q and Wigner functions of
$|\psi\rangle_\alpha $ are simply a displacement
of those of $|\psi\rangle$, namely
\begin{equation}
    Q(\beta)_{|\psi\rangle_\alpha}=
    Q(\beta-\alpha)_{|\psi\rangle},\ \ \ \
    W(\beta)_{|\psi\rangle_\alpha}=
    W(\beta-\alpha)_{|\psi\rangle}. \label{QWIG}
\end{equation}
So the Q-function and the Wigner
function of the state $\|\eta,M\rangle $ are easily obtained
from those of the photon-added coherent states given in \cite{Aga}
\begin{eqnarray}
    Q(\beta)&=&|\langle \beta\|\eta,M\rangle|^2=
         \frac{e^{-|\beta|^2}|\lambda+\beta|^{2M}}{M!L_M(-\lambda^2)},
         \label{qfunc}\\
    W(\beta)&=&\frac{2(-1)^M L_M\left(|2\beta+\lambda|^2
    \right)}{\pi L_M(-\lambda^2)} \exp\left(-2|\beta|^2\right).
\end{eqnarray}

The Q-function Eq.(\ref{qfunc}) has a $2M$-fold zero at the
position $\beta = -\lambda$, which signals the nonclassical behaviour\footnote{We thank the referee for this remark.}.
These zeros are  related to the negative parts of the
Wigner function, since the Q-function can be defined as a
smoothed Wigner function.  Fig.\,4 gives plots of the Wigner
function of $\|\eta, M\rangle $ for $ M=3$ and different $\eta $.
One can clearily see the negative parts, except for the
case $\eta=0$ which corresponds to the vacuum state whose Wigner function is simply a Gaussian centered at the origin.
As $\eta$ increases from 0, the Gaussian distribution continuously
deforms to the Wigner function of the number state
$|3\rangle $.

 We can also study squeezing
properties from the Q-function by examining the deformation of its
contours. Fig.5 is the contour plot of Q functions
for $ M=10 $ and different $\eta $. We see that, when
we increase $\eta $, the contour is squeezed
in the $ x$ direction until a maximum squeezing
is reached. Then the contour deforms to the shape
of a banana, which occupies a wider range in the
$ x $ direction and the squeezing is reduced. Finally,
we obtain a circular contour for larger $\eta $
corresponding to no squeezing (c.f. Fig.\,2).

\section{Generation of intermediate states}\label{intn}
\setcounter{equation}{0}

The main difference between the intermediate  states described herein and  photon-added
coherent states is that the former  are a {\em finite} superposition of
number states.  This suggests the possibility of
an experiment to produce
 these states using the method proposed in  \cite{jdsa}.

We can also generate the state $\|\eta,M\rangle $ by using the interaction
of a photon and a two-level atom with an external
classical driving
field $A$ in a cavity.  In the rotating wave approximation,
the Hamiltonian($ \hbar=1 $) is
\begin{eqnarray}
     && H=H_0+V,  \nonumber \\
     && H_0=\omega N + A(a^\dagger + a) +
     \frac{1}{2}\omega_0 \sigma_3,
     \nonumber \\
     && V=g(a^\dagger \sigma_- +a\sigma_+),
     \label{Hamil1}
\end{eqnarray}
where $ \sigma_3=|e\rangle\langle e|-|g\rangle\langle g|$,
$ \sigma_+=|e\rangle\langle g| $ and $ \sigma_-=|g\rangle\langle e|$
are atomic operators, $ g $ is the one-photon coupling constant, $\omega
_{0}$ and $\omega $ are the atomic transition frequency and cavity resonant
mode frequency respectively, and we take the driving field $A$ to be  real and
constant.
The interaction Hamiltonian
is
\begin{equation}
    H_I (t) = U_0^{-1}(t) V U_0(t), \ \ \ \ \
    U_0(t) = e^{-iH_0 t}=e^{-{i/2}\omega t N -iAt (a^\dagger +a)}
                   e^{-i\omega_0 t \sigma_3}.
\end{equation}
Using the following relation (see Appendix B)
\begin{equation}
    U_0^{-1}(t) a U_0(t)=e^{-i\omega t} D(-A/\omega) a D(A/\omega),
    \label{formuaaa}
\end{equation}
where $ D(A/\omega) $ is the displacement operator,
we have
\begin{equation}
   H_I(t)=g D(-A/\omega)\left(e^{i(\omega-\omega_0)t}a^\dagger \sigma_-
   +e^{-i(\omega-\omega_0)t}a \sigma_+ \right) D(A/\omega).
\end{equation}
Now we consider the on-resonance case, $\omega=\omega_0$. Then the
interaction Hamiltonian is time-independent
\begin{equation}
   H_I = g D(-A/\omega) (a^\dagger \sigma_- + a \sigma_+)D(A/\omega)
\end{equation}
and therefore its time evolution operator is
\begin{equation}
   U_I(t)=e^{-iH_I t}=D(-A/\omega)
   e^{-igt(a^\dagger \sigma_- + a \sigma_+)} D(A/\omega).
\end{equation}
Suppose that the field is initially prepared in the vacuum state $|0\rangle$
and the atom in the excited state $|e\rangle $; namely, at $t=0$, the system
is in the state $ |0\rangle \otimes |e\rangle $. At time $ t$ we have
\begin{equation}
   U_I(t)  |0\rangle \otimes |e\rangle =
   D(-A/\omega)e^{-igt(a^\dagger \sigma_- + a \sigma_+)} D(A/\omega)
   |0\rangle \otimes |e\rangle.
\end{equation}
When $ gt \ll  1 $, we have
\begin{eqnarray}
  U_I(t)  |0\rangle \otimes |e\rangle &=& |0\rangle \otimes |e\rangle
  -igt \left[ D(-A/\omega) a^\dagger D(A/\omega)|0\rangle \right]
  \otimes |g\rangle.
\end{eqnarray}
If the atom is detected in the ground state $ |g\rangle $,  the field
is reduced to the state $ \|\eta, 1 \rangle $ with
$ \eta=\omega^2/(A^2+\omega^2) $.

The state $ \|\eta,M\rangle $ ($ M>1 $) can be generated by a
multiphoton generalization of the HamiltonianEq.(\ref{Hamil1}),
namely, $ V=g(a^{\dagger M} \sigma_- +a^M \sigma_+) $.

Note that the parameter $A$ depends on the external
driving field and is a tunable parameter. In particular, for large
enough $ M $, we can control the output state to be either a number or
a coherent state by tuning the parameter $ A $.

Finally we may infer the presence of these new intermediate
states to first order in an idealized non-linear optics experiment.
Consider a nonlinear Mach-Zehnder interferometer
with a Kerr medium in one arm. The output state is
the displaced Kerr state \cite{Prod}
\begin{equation}
    D(\xi)U_K(\gamma) |\lambda\rangle, \ \ \ \ \
    U_K(\gamma)\equiv \exp\left(\frac{i}{2}\gamma
    a^{\dagger 2}a^2\right),
\end{equation}
where $D(\xi)$ is the displacement operator and
$\gamma\equiv 2\chi L/v$,  $L$ is the length of
the Kerr medium,  $ v $ the appropriate phase
velocity inside the medium and $\chi$ the third-order
susceptibility.  When $\xi=-\lambda$, and $\gamma $
is small enough, the above states can be approximated as
\begin{equation}
    |0\rangle+\frac{i}{2}\gamma\lambda^2 \|\lambda,2\rangle
\end{equation}
showing the presence of the state $|\eta,2\rangle$ in first order.
In general, if we use a $(2S+1)$th-order nonlinear Kerr medium
modelled in the interaction picture by \cite{splus1}
\begin{equation}
    H_{\mbox{\scriptsize Kerr}}=
    \frac{\hbar \gamma_S}{(S+1)!} (a^\dagger)^{S+1}
    a^{(S+1)} = \frac{\hbar \gamma_S}{(S+1)!}
    N(N-1)\cdots (N-S),
\end{equation}
we can find $ \|\eta,M\rangle $ when $ \gamma_S $ is small enough.

\section{Interaction with a two-level atomic system}
\setcounter{equation}{0}

In this section we turn to the interaction of the state
$\|\eta,M\rangle $ with a simple two-level system in the framework
of the two-photon Jaynes-Cummings model (JCM) \cite{jay,sho}
described by the following Hamiltonian ($\hbar =1$)
\begin{equation}
H=\omega a^{\dagger }a+\frac{1}{2}\omega _{0}\sigma _{3}+
g(a^{\dagger 2}\sigma_{-} + a^{2} \sigma_{+})=H_{0}+V
\end{equation}
with
\[
H_{0} =\omega a^{\dagger }a+\frac{1}{2}\omega _{0}\sigma _{3},
\ \ \ \ \
V =g(a^{\dagger 2}\sigma_{-} + a^{2} \sigma_{+}).
\]
The notation is as in Eq.\,(\ref{Hamil1}), but now  $g$ is the two-photon coupling
constant for transition $|g\rangle \rightleftharpoons |e\rangle $.
Suppose that, at the initial time $t=0$, atom and field are
decoupled and the atom is initially in the excited state $|e\rangle $, while
the field is in the intermediate number-coherent state $||\eta ,M\rangle $.
Then the combined atom-field wave function at time $t$ is obtained
as
\begin{eqnarray}
|\psi _{I}(t)\rangle  &=&\sum_{n=0}^{M}C_{n}(\eta ,M)[\cos (\delta _{n}t)-i%
\frac{\Delta }{2\delta }\sin (\delta _{n}t)]e^{i\frac{\Delta }{2}t}|e\rangle
\otimes |n\rangle   \nonumber \\
&&-i\sum_{n=0}^{M}\frac{\Omega _{n}}{\delta _{n}}C_{n}(\eta ,M)\sin (\delta
_{n}t)e^{-i\frac{\Delta }{2}t}|g\rangle \otimes |n+2\rangle ,
\label{solution}
\end{eqnarray}
where
\begin{equation}
    \Omega_{n} = g\sqrt{(n+1)(n+2)}, \ \ \ \ \
    \delta_{n} =\sqrt{\frac{\Delta^{2}}{4}+\Omega _{n}^{2}},
    \ \ \ \ \
    \Delta =\omega_{0}-2\omega.
\end{equation}
For simplicity, we only consider the on-resonance interaction case
$\Delta =0$ as in \cite{buzek} whereupon  Eq.(\ref{solution}) simplifies to
\begin{equation}
|\psi _{I}(t)\rangle =\sum_{n=0}^{M}C_{n}(\eta ,M)\cos (\Omega
_{n}t)|e\rangle \otimes |n\rangle -i\sum_{n=0}^{M}C_{n}(\eta ,M)\sin (\Omega
_{n}t)|g\rangle \otimes |n+2\rangle.  \label{solution1}
\end{equation}
We now discuss some quantum characteristics of the system arising
from the equation Eq.(\ref{solution1}).

\subsection{Atomic population inversion}

Atomic population inversion is an
important atomic observable in the JCM and is defined as the difference
between the probabilities of finding the atom in the excited state
and in the ground state. From Eq.(\ref{solution1}), the atomic
population inversion is obtained as
\begin{equation}
     W(t)=\langle \sigma _{3}\rangle =
     \sum_{n=0}^{M}|C_{n}(\eta ,M)|^{2}
     \cos(2\Omega _{n}t). \label{pop}
\end{equation}

Fig.\,6 gives the inversion vs. scaled time $\tau\equiv gt$ for
different $M$ and $\eta $. From Fig.\,6, we observe
that the atomic population inversion exhibits the conventional
Rabi oscillation for the $M$-number state limit ($\eta\to 1$).
In fact, in the limit $\eta\to 1$,  Eq.\,(\ref{pop}) is simplified
as
\begin{equation}
   W(t)=\cos(2\Omega_M t)
\end{equation}
with frequency $2\Omega_{M}=2g[(M+1)(M+2)]^{1/2}$
($\approx 2Mg$ for high enough $\langle N\rangle $, see Fig.6(a)).
In the coherent state limit we observe the collapse-revival phenomenon,
as we expect, with a revival time $t_{\mbox{\scriptsize cs}}$ which can be
estimated as $\pi /g$ \cite{buzek} for high enough
$\langle N \rangle $ (that is, {\em revival frequency}
$\Omega_{\mbox{\scriptsize cs}}\equiv 2\pi/t_{\mbox{\scriptsize cs}}
\approx 2g $) (Fig.6(d)). For the
general intermediate case (Fig.6(b,\,c)), remnants of both
behaviour are seen; namely, an oscillation of frequency $\Omega
_{M}$ modulated by the frequency $\Omega_{\mbox{\scriptsize cs}}$
with modulated
amplitude dependent on the parameter $\eta $ and $M$.

\subsection{Field entropy}

We now consider the cavity field observables, beginning with
 entropy which is a measure of the {\em amount
of chaos} or lack of information about a system \cite{alfred}.
The entropy $S$ of a quantum-mechanical system is defined
as \cite{barnett,buzek}
\[
S=-\mbox{Tr}(\rho \ln (\rho )),
\]
where $\rho $ is the density operator of the quantum system
and the Boltzmann constant $k$ is  equal to unity.
For a pure state, $S=0$; otherwise $S>0$, and it increases
with increasing number of microstates with decreasing
statistical weight.

In this subsection we study the time evolution of the field
entropy in our system. Barnett and Phoenix \cite{barnett}
have proved that the field entropy $S_{f}$ equals the atomic
entropy $S_{a}$ if the total initial state is a
pure state. From Eq.(\ref{solution1}) the atomic reduced
density operator $\rho_{a}$ can be easily obtained as
\begin{equation}
     \rho _{a}\equiv Tr_{f}(\rho)=\rho _{11}|g\rangle
     \langle g|+\rho _{12}|g\rangle \langle e|+\rho_{21}
     |e\rangle \langle g|+\rho _{22}|e\rangle \langle e|,
\end{equation}
where
\begin{eqnarray}
     && \rho _{11}=\sum_{n=0}^{M}|C_{n}(\eta ,M)|^{2}
                            \sin ^{2}(\Omega _{n}t) \nonumber \\
     &&\rho _{22} =\sum_{n=0}^{M}|C_{n}(\eta ,M)|^{2}
                            \cos ^{2}(\Omega _{n}t) \\
     &&\rho _{12} =\rho _{21}^{\ast }=\sum_{n=0}^{M-2}
         C_{n+2}(\eta ,M)C_{n}(\eta,M)
         \cos (\Omega _{n+2}t)\sin (\Omega _{n}t).  \nonumber
\end{eqnarray}
Then the field and atomic entropy
$S_{a}=-\mbox{Tr}_{a}(\rho _{a}\ln (\rho _{a})) $
can be expressed as
\begin{equation}
     S_{f}=S_{a}=-\pi _{+}\ln (\pi _{+})-\pi _{-}\ln (\pi _{-}),
\end{equation}
where $\pi _{\pm}$ are eigenvalues of the atomic reduced field
density operator $\rho _{a}$
\begin{equation}
    \pi _{\pm }=\frac{1}{2}(1\pm \sqrt{(\rho _{22}-
    \rho _{11})^{2}+4| \rho_{12}|^{2}}).
\end{equation}

The field entropy $S_{f}$ as a function of $\tau$
is presented in Fig.\,7.
It is clear that $S_{f}$ is a periodic function of time and it exhibits the
conventional oscillation for the $M$-number state limit. As in the case
of coherent initial states, the field entropy during the time evolution is
dynamically reduced to zero at revival time $t_R$ which means the cavity
field can be periodically found in pure states, and reaches a maximum at
$t_R/2$ and falls quickly to a minimum at $ \tau=\pi/4, 3\pi/4$.
Furthermore, for the general intermediate case, the field entropy
has more minima as shown in Fig.\,7 (b,\,c) due to the frequency
modulation.

\subsection{Q-function}

The quasi-probability distribution Q-function is defined as \cite{Q-fun}:
\[
     Q(\beta )=\frac{1}{\pi }\langle \beta |\rho_f |\beta \rangle,
\]
where $\rho_f = \mbox{Tr}_{a}(\rho)$ is the field reduced density operator
\begin{eqnarray}
     \rho _{f}
     &=&\sum_{m,n=0}^{M}C_{m}(\eta ,M)C_{n}(\eta ,M)
     \left[\cos (\Omega _{m}t)\cos
     (\Omega _{n}t)|n\rangle \langle m| \right. \nonumber \\
    & & \left.+\sin (\Omega _{m}t)\sin (\Omega _{n}t)
    |n+2\rangle \langle m+2|\right], \label{rdmf}
\end{eqnarray}
and $|\beta\rangle$ is the coherent state.
So the Q-function of the cavity field is
\begin{equation}
Q(\beta )=\frac{e^{-|\beta |^{2}}}{\pi }\left( \left| \sum_{n=0}^{M}\frac{
\beta^{\ast n}}{\sqrt{n!}}C_{n}(\eta ,M)\cos (\Omega _{n}t)\right|
^{2}+\left| \sum_{n=0}^{M}\frac{(\beta^{\ast })^{n+2}}{\sqrt{(n+2)!}}
C_{n}(\eta ,M)\sin (\Omega_{n}t)\right|^{2}\right) .
\end{equation}

In Fig.\,8 we give contour plots of the Q-function at different times
$\tau $ for $\eta=0.1,\ 0.8$. At time $\tau =0$, the Q-function has only a
single peak and the field is in the pure quantum state $|\eta,M\rangle$
(c.f. Fig.\,5). With the development of time, the Q-function
begins to separate into two peaks. The smaller $\eta $, the faster
the seperation. At time $\tau =\pi/2$, the Q-function exhibits the most
separation and the field is in a mixed state since the entropy
reaches its maximum. Then two peaks begins to merge together
and finally combine in a single peak at time $\tau=\pi$,
where the field is in a pure state with vanishing entropy.

\subsection{Photon number distribution}

The photon number distribution $P_{n}(t)$ of the field described
by the reduced density matrix $\rho_f $ is given by
\begin{equation}
P_{n}(t)=\langle n|\rho _{f}|n\rangle.
\label{phdis}
\end{equation}
Inserting Eq.(\ref{rdmf}) into Eq.(\ref{phdis}) we find
 the photon number distribution at time $t$
\begin{equation}
P_{n}(t)=|C_{n}(\eta ,M)|^{2}\cos ^{2}(\Omega _{n}t)+|C_{n-2}(\eta
,M)|^{2}\sin ^{2}(\Omega _{n-2}t).  \label{PND}
\end{equation}

Fig.\,9 shows the behaviour of the photon number distribution
at times $\tau=0, \pi /4$, $\pi /2$, $3\pi/4$ and $\pi $. From these figures
we can observe that the photon number distribution exhibits
strong oscillation at time $\tau =\pi/4$ and $ 3/4\pi $ for
the intermediate states.
In fact, at those times, the field is a superposition of two
components (see Fig.8) and its emtropy decreases rapidly 
to a minimum (see Fig.7).  Partial interference between
two component results in strong oscillation of the photon
number distribution. However, the oscillation is not {\em perfect} (see below).
Nevertheless it is perfect at the slightly earlier time
$\tau=\pi/4-\xi $ (see dashed lines in Fig.9(a,b)).

This effect is not hard to understand. In fact, at $\tau=\pi/4$,
we have the following approximate result (see Apendix C)
\begin{equation}
  P_n(t)=\left[1+\frac{(1-\eta)^2 n(n-1)}
  {\eta^2(M-n+2)^2(M-n+1)^2}\right]|C_n(\eta,M)|^2
  \sin^2\left[\left(n-\frac{1}{2}\right)\tau
  \right]_{\tau=\frac{\pi}{4}} \label{614}
\end{equation}
for the high enough $\langle N\rangle $ case.
Eq.(\ref{614}) is a strongly oscillating function which explains
the large oscillations of the photon number distribution. However,
due to the additional term $\tau/2=\pi/8$, the function
$\sin^2[(n-1/2)\tau]_{\tau=\pi/4}$ cannot be zero for any
integer $n$; in other words, the oscillation is not perfect.
However, $P_n(t)$   is zero at the slightly  earlier time $\tau=\pi/4-\xi$,
where $\xi$ is chosen to make $(n-1/2)\tau$ a multiple of $\pi$.

From Fig.9(c) we also observe that the photon number
distribution at $\tau=\pi$ is simply a displacement by 2 from that
at the time $\tau=0$. For the large photon number case, this
fact can be proved analytically. Using Eq.\,(\ref{c444}) in
Appendix C, we have
\begin{eqnarray}
   &&\sin(\Omega_{n-2}t)\approx
        \sin(n\pi-\pi/2)=(-1)^{n+1}, \nonumber \\
   &&\cos(\Omega_n t)\approx
        \cos[(n+1)\pi+\pi/2]=0. \nonumber
\end{eqnarray}
So the photon number distribution Eq.(\ref{PND}) at $\tau=\pi$ becomes
\begin{equation}
    P_n(\pi/g)=|C_{n-2}(\eta,M)|^2 \equiv P_{n-2}(0).
\label{lasteqo}
\end{equation}

In the same way we find that, at $\tau=\pi/2$, the photon
number distribution is
\begin{equation}
    P_{n}(\pi/2g)=\frac{1}{2}\left(|C_{n}(\eta ,M)|^{2}+
    |C_{n-2}(\eta, M)|^{2}\right) =
    \frac{1}{2} \left( P_n(0)+P_n(\pi/g)\right),  \label{lasteq}
\end{equation}
namely, the {\em average}  of the photon number distributions
at $\tau=0$ and $\tau=\pi$.  In Fig.9(c) this fact can be
clearly observed.

\section{Conclusion}

In this paper we have  described new states $\|\eta,M\rangle$
which interpolate between number and coherent states and have
investigated their various properties. Unlike photon-added
coherent states, to which they are related, these states are a
{\em finite} superposition of number states. We also analyzed
the limiting cases  $\eta \rightarrow 1$ and
$\eta \rightarrow 0,$ $M\rightarrow \infty $ corresponding to
number and  coherent states respectively.
Salient statistical properties of $\|\eta,M\rangle $
 such as the sub-Poissonian distribution, the
 anti-bunching effect and the
squeezing effects were investigated for a wide range
of parameters. The non-classical features of thse states
for certain parameter ranges were
demonstrated in terms of the quasiprobability
distributions, the Q and Wigner functions.
We also proposed an experiment to generate these states,
inferring their presence in certain non-linear systems.

We then considered the interaction of these
interpolating number-coherent states with a two-level
atomic system, exemplified by the two-photon Jaynes-Cummings
Model.  We first studied the dynamics of atomic population
inversion. On an intuitive level, one expects that the response
of the atomic system will vary between the Rabi oscillation
typical of an initial number state, and the collapse-revival
mode for an initial coherent state; and indeed this is what
one obtains. We found that it exhibited the
conventional Rabi oscillation for the $M$-number state limit
with frequency $\Omega _{M}$ ($\approx 2Mg$) and the
collapse-revival phenomenon for the coherent state limit
with rivival frequency $\Omega_{\mbox{\scriptsize cs}}
\approx 2g $. For the general intermediate case,
remnants of both behavior were seen; namely, an oscillation
of frequency $\Omega_{M}$ modulated by the frequency
$\Omega_{\mbox{\scriptsize cs}}$ with modulated amplitude dependent
on the parameter $\eta $ and $M$.

We further investigated the field observables, the entropy,
Q-function and photon number distribution. It is of interest
that the photon number distribution exhibit strong oscillation
at $\tau=\pi/4, 3\pi/4 $. At those times, the field
entropy falls rapidly to a minimum and the Q-function
separates into two peaks, which means that the field is a
superposition of almost pure states and interference between
components of the superposition state leads to strong oscillation
of the photon number distribution. An approximate analytical
solution is presented to explain this result.

The remarkable properties of these intermediate number-coherent
states provide a useful tool for theoretical investigation of model
systems; their generation by  non-linear systems tempts us to believe
that the states found in this paper may play an important role in
quantum optics.

\section*{Acknowledgements}

H.\,Fu is supported in part by the National Natural Science
Foundation of China.  We would like to thank Dr.A.Greentree
of the Quantum Processes Group of the Open University for
enlightening discussion on the production of these states, and
the referee for valuable comments.

\section*{Appendix A: Displaced quasiprobability distributions}
\setcounter{equation}{0}
\renewcommand{\theequation}{A.\arabic{equation}}
For the Q-function, we prove Eq.\,(\ref{QWIG}) as follows:
\begin{eqnarray}
    Q(\beta)_{|\psi\rangle_\alpha}&=&
    |\langle\beta|D(\alpha)|\psi\rangle|^2=
    |\langle 0|D(-\beta)D(\alpha)|\psi\rangle|^2 \nonumber\\
    &=& |\langle \beta-\alpha|\psi\rangle|^2=
    Q(\beta-\alpha)_{|\psi\rangle},
\end{eqnarray}
where we have used the relation
\begin{equation}
    D(\delta)D(\gamma)=D(\delta +\gamma)e^{\frac{1}{2}
    (\delta\gamma^*-\gamma\delta^*)}
    =D(\delta+\gamma)e^{i\mbox{\scriptsize Im}(\delta\gamma^*)},
\end{equation}
for arbitrary complex numbers $\delta $ and $\gamma $. From the
following definition of the Wigner function \cite{wig}
\begin{equation}
   W(\beta)=\frac{2}{\pi}\sum_{k=0}^\infty
             \langle\beta,k|\rho|\beta,k\rangle,
   \label{wig1}
\end{equation}
where $|\beta,k\rangle\equiv D(\beta)|k\rangle=
e^{\beta a^\dagger -\beta^* a}|k\rangle $ is the displaced
number state ($|k\rangle$ is the number state) and
$\rho=|\eta,M\rangle\langle \eta,M|$
is the density matrix of the states considered, we can prove
the second relation in Eq.(\ref{QWIG}) in the same way as
in the Q-function case.

\section*{Appendix B: Proof of formula Eq.(\ref{formuaaa})}
\setcounter{equation}{0}
\renewcommand{\theequation}{B.\arabic{equation}}

In this appendix we give a proof of Eq.(\ref{formuaaa}). We use the
following formula
\begin{equation}
   e^{-F}G e^{F}=\sum_{n=0}^\infty \frac{(-1)^n}{n!}
   \underbrace{[F, [F, \cdots, [F}_{n \mbox{\scriptsize\ copies}},G]\cdots]].
   \label{formu}
 \end{equation}
For the case in hand
\begin{equation}
   F=-i\omega t N-itA(a^\dagger + a), \ \ \ \ \
   G=a.
\end{equation}
It is easy to see that
\begin{eqnarray}
   && [F,G]=i\omega t a +iAt, \nonumber \\
   && [F,[F,G]]=i\omega t [F, G], \nonumber \\
   && [F,[F,[F,G]]]=i\omega t [F,[F,G]]=(i\omega t)^2 [F,G], \nonumber \\
   && \cdots \cdots \nonumber \\
   && \underbrace{[F,[F, \cdots, [F}_{n \mbox{\scriptsize \ copies}},
   G]\cdots ]]=(i\omega t)^n [F,G]\nonumber \\
   &&=(i\omega t)^n a +(i\omega t)^n A/\omega
   =(i\omega t)^n D(-A/\omega)aD(A/\omega), \label{aaapp}
\end{eqnarray}
where $ D(A/\omega) $ is the displaced operator.
Substituting Eq.\,(\ref{aaapp}) into Eq.\,(\ref{formu}) we obtain the
formula Eq.(\ref{formuaaa}).

\section*{Appendix C: Photon number distribution for large photon number}
\setcounter{equation}{0}
\renewcommand{\theequation}{C.\arabic{equation}}

In this appendix we present an analytical treatment of the photon number
distribution in the large photon number regime.  The photon number distribution
of the two-photon JC model with initial state
$|e\rangle \otimes \sum_n C_n|n\rangle $ can be obtained as
\begin{equation}
  P_n(t)=|C_n|^2 \cos^2\left(\sqrt{(n+1)(n+2)}\tau\right)+
  |C_{n-2}|^2 \sin^2\left(\sqrt{(n-1)n}\tau\right),\label{c1}
\end{equation}
where $\tau=gt$ is the scaled time as before.

Here we only consider an initial field state which is narrower than that of a coherent state. For a distribution $\{ |C_n|^2\}$ we can calculate
the variance as 
\begin{equation}
  \left(n-\bar{n}\right)^2 =
  \langle N^2\rangle -\langle N\rangle^2.
\end{equation}
For the coherent state $|\alpha\rangle$, we have 
$ \left(n-\bar{n}\right)^2 = \bar{n}$. So for highly excited coherent
states  where $\bar{n}\to \infty $, we have $n\sim \bar{n}$.  In the
following we only consider a distribution $\{|C_n|^2\}$
 narrower than  the Poisson distribution, namely
\begin{equation}
  \left(n-\bar{n}\right)^2 \leq \bar{n}.
\end{equation}
So for large enough $\bar{n}$ we also have $ n\sim \bar{n}$. In
this case, we have
\begin{eqnarray}
  &&\sqrt{(n+1)(n+2)}\approx \sqrt{n^2 +3n}=
          n\sqrt{1+\frac{3}{n}}=
     n+\frac{3}{2}=\left(n-\frac{1}{2}\right)+2,
     \nonumber \\
  &&\sqrt{(n-1)n}\approx \sqrt{n^2 -n}=
          n\sqrt{1-\frac{1}{n}}= n-\frac{1}{2}. \label{c444}
\end{eqnarray}
Futhermore, when $\tau=\pi/4 $ we have
\begin{equation}
  \cos^2\left(\sqrt{(n+1)(n+2)}\tau\right)=
  \cos^2\left[\left(n-\frac{1}{2}\right)\tau+\frac{\pi}{2}
  \right]_{\tau=\frac{\pi}{4}}
  =\sin^2\left[\left(n-\frac{1}{2}\right)\tau\right]_{
  \tau=\frac{\pi}{4}}. \label{c5}
\end{equation}
Substituting Eq.(\ref{c5}) into Eq.(\ref{c1}) we obtain the approximate photon number distribution at $\tau=\pi/4$
\begin{equation}
  P_n(t)=\left(|C_n|^2 +|C_{n-2}|^2\right)
  \sin^2\left[\left(n-\frac{1}{2}\right)\tau\right]_{
  \tau=\frac{\pi}{4}}.\label{9}
\end{equation}
From Eq.(\ref{9}) we find that, for the initial field whose photon
distribution is narrower than a Poisson distribution, the photon number
distribution at $\tau=\pi/4$ exhibits strong oscillation. However,
$P_n(\pi/4)$ cannot be zero for any $n$ due to the term $\tau/2=
\pi/8$ and the oscillation is not perfect. Nevertheless, the oscillation
is perfect at a slightly earlier time $\tau=\pi/4-\xi $, as indicated in
Ref.\cite{buzek} (for initial coherent state) and Fig.9(a,b).

For coherent states we further have $ |C_{n-2}|^2\approx |C_{n}|^2 $.
So the photon number distribution is
\begin{equation}
  P_n(t)=2 e^{-\bar{n}}\frac{\bar{n}^n}{n!}
  \sin^2\left[\left(n-\frac{1}{2}\right)\tau\right]_{
  \tau=\frac{\pi}{4}},
\end{equation}
which is just the result given in Ref.\,\cite{buzek}.

Now we turn to the analytical approximate result Eq.(\ref{614}).
For the intermediate state, we can write the variance as
\begin{equation}
  (n-\bar{n}_M)^2 =\langle N\rangle_{M-1}
  \langle N\rangle_M -\langle N\rangle_M^2 +
  \langle N\rangle_M,
\end{equation}
where $\bar{n}_M\equiv \langle N\rangle_M\equiv
\langle \eta,M|N|\eta,M\rangle$.
In general, we have $\langle N\rangle_{M-1}
\leq \langle N\rangle_M\leq M $. For large enough
$\bar{n}_M$, or  $M$,
we have $\langle N\rangle_{M-1}\approx\langle N\rangle_M$
and therefore $(n-\bar{n}_M)^2\sim \bar{n}_M$ which leads to
$n \sim \bar{n}_M$.  So the result Eq.(\ref{9}) is valid for the
intermediate state case. Furthermore, the
distribution $|C_n(\eta,M)|^2$ and $|C_{n-2}(\eta,M)|^2$
are related by
\begin{equation}
  |C_{n-2}(\eta,M)|^2=\frac{(1-\eta)^2 n(n-1)
  }{\eta^2 (M-n+2)^2 (M-n+1)^2} |C_n(\eta,M)|^2.
  \label{hello}
\end{equation}
Sunstituting Eq.(\ref{hello}) into Eq.(\ref{9}), we finally obtain
Eq.(\ref{614}).

Eq.(\ref{c444}) can also be used to explain the behaviour of
the photon number distribution at $\tau=\pi/2$ and $\pi$
(see Eqs.(\ref{lasteqo},\ref{lasteq}) and Fig.9).


\newpage

\input{epsf}

\begin{figure}
\centerline{\epsfxsize=8cm
\epsfbox{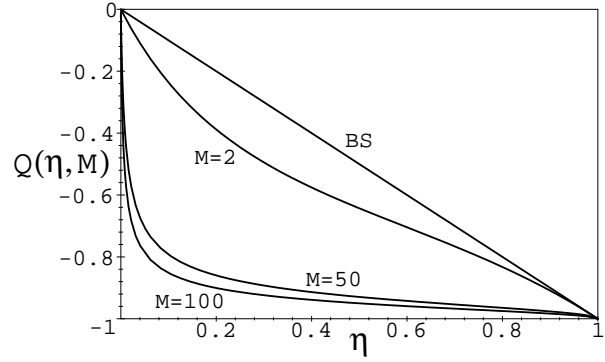}}
\caption{Mandel's Q parameter for $M$=2, 50, 100. }
\end{figure}


\begin{figure}
\centerline{\epsfxsize=8cm
\epsfbox{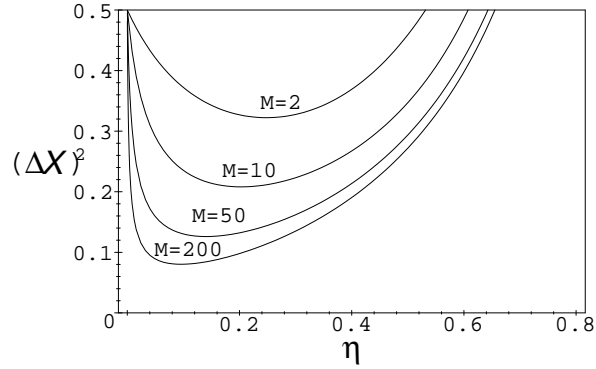}}
\caption{Variance $(\Delta x)^2 $ of $\|\eta,M\rangle $
as a function of $\eta $ for $M=2, 20, 50$,and 200.}
\end{figure}


\begin{figure}
\centerline{\epsfxsize=8cm
\epsfbox{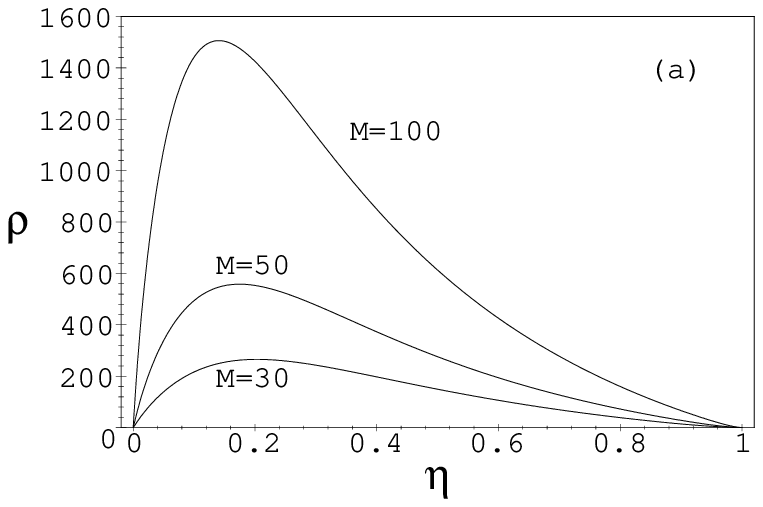}
\epsfxsize=8cm
\epsfbox{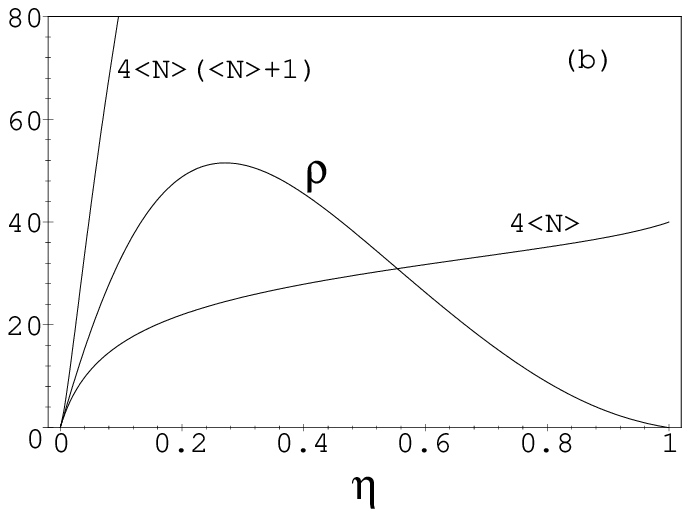}}
\caption{The signal-to-quantum noise ratio for $\|\eta,M\rangle $:
(a) The ratio for different $ M$; (b) Comparison of $\rho$ ,
$ 4\langle N\rangle (\langle N\rangle +1)$ and
$ 4\langle N\rangle$ for $M=10$.}
\end{figure}

\newpage


\begin{figure}
\centerline{\epsfxsize=8cm
\epsfbox{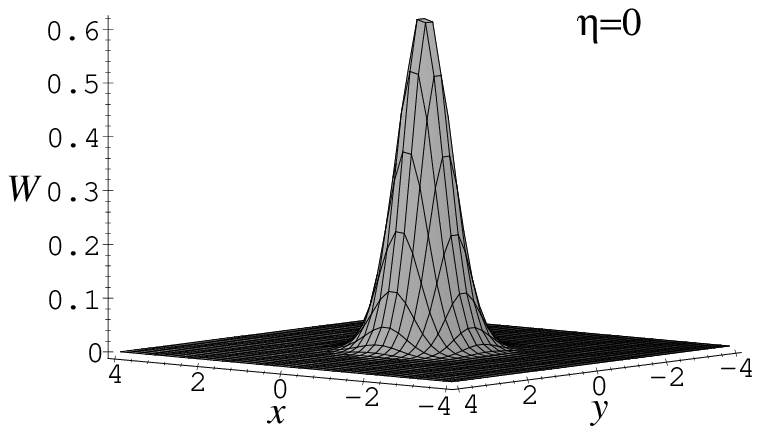}
\epsfxsize=8cm
\epsfbox{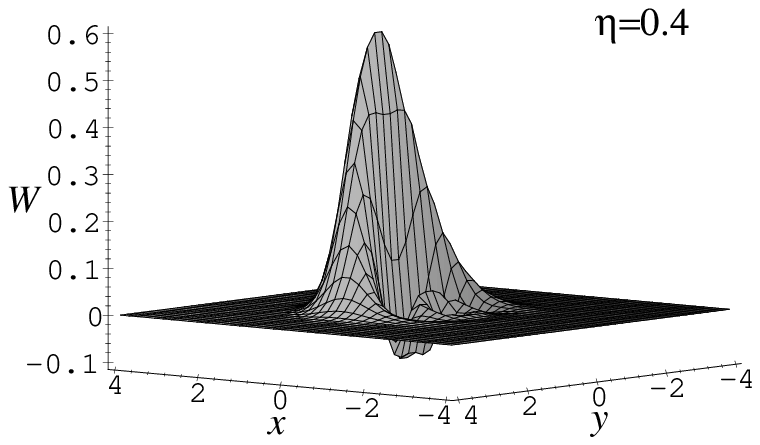}}
\centerline{\epsfxsize=8cm
\epsfbox{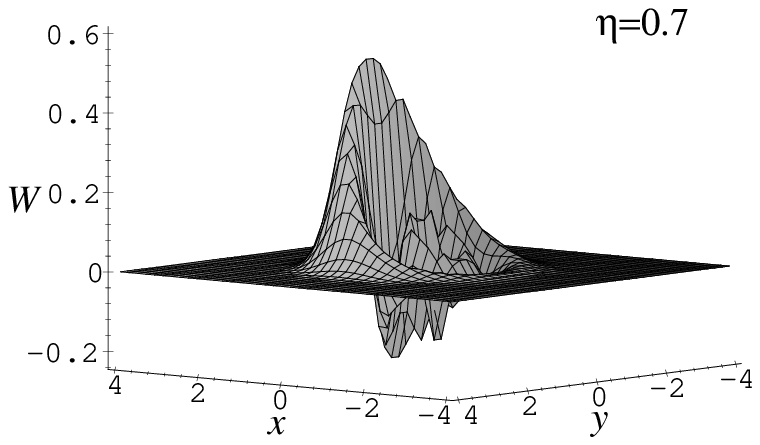}
\epsfxsize=8cm
\epsfbox{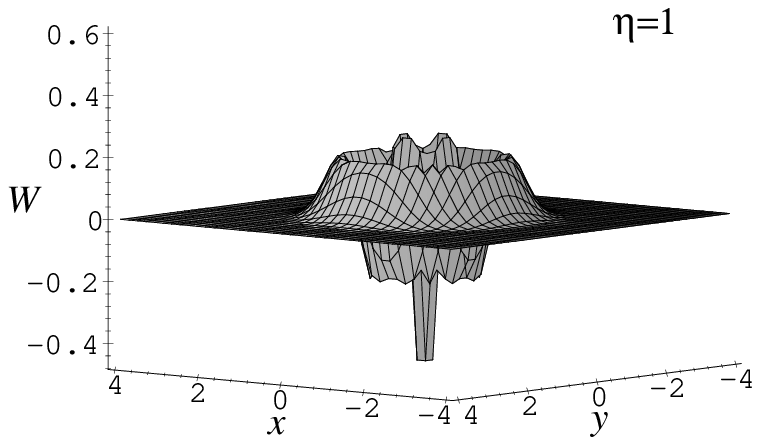}}
\caption{Wigner function of $\|\eta,M\rangle $ for $M=3$
and $\eta=0.1,0.4,0.7$ and 1. $\alpha=x+iy$. }
\end{figure}

\newpage

\begin{figure}
\centerline{\epsfxsize=7cm
\epsfbox{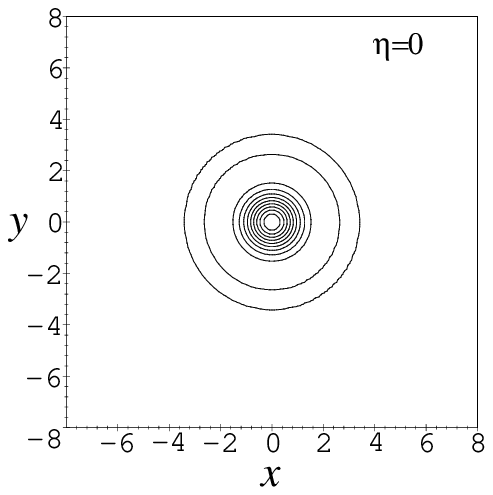}
\epsfxsize=7cm
\epsfbox{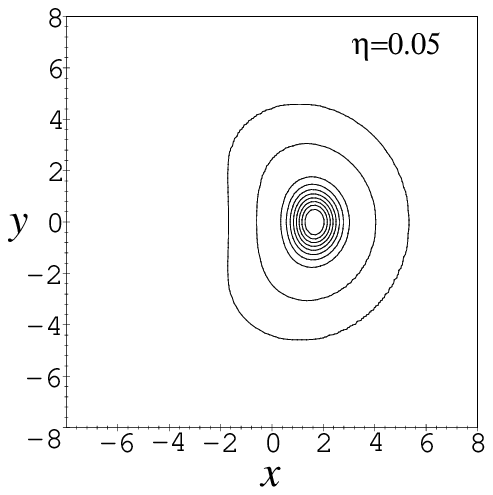}}
\centerline{\epsfxsize=7cm
\epsfbox{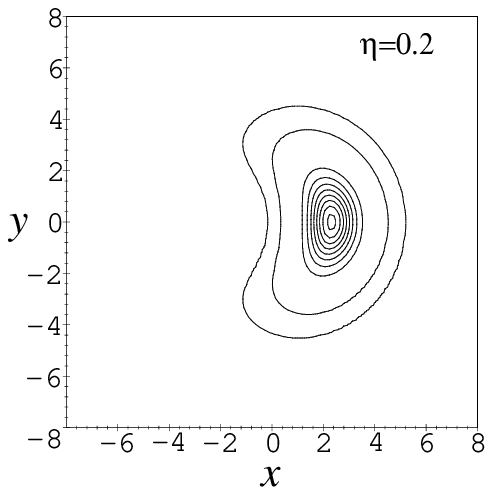}
\epsfxsize=7cm
\epsfbox{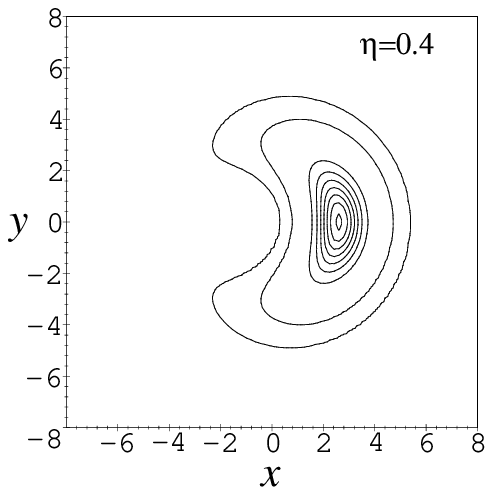}}
\centerline{\epsfxsize=7cm
\epsfbox{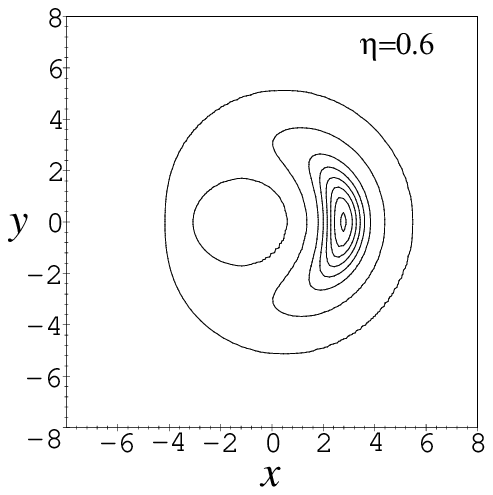}
\epsfxsize=7cm
\epsfbox{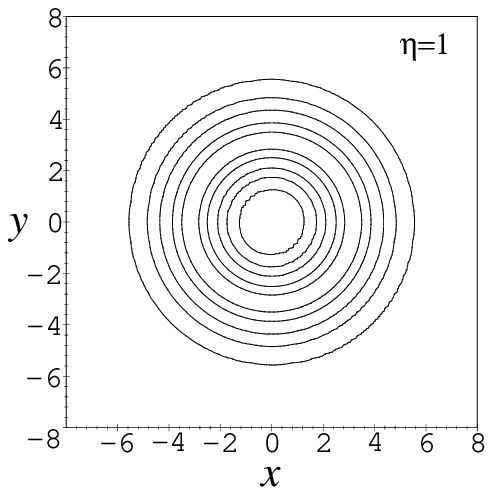}}
\caption{Contours of the Q-function of $\|\eta,M\rangle$.
In all cases $ M=10 $. $\alpha=x+iy$.  }
\end{figure}


\newpage

\begin{figure}
\centerline{\epsfxsize=8cm
\epsfbox{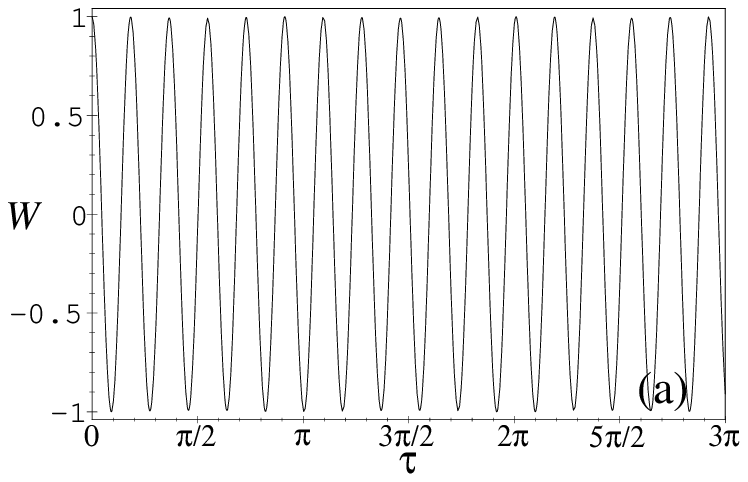}
\epsfxsize=8cm
\epsfbox{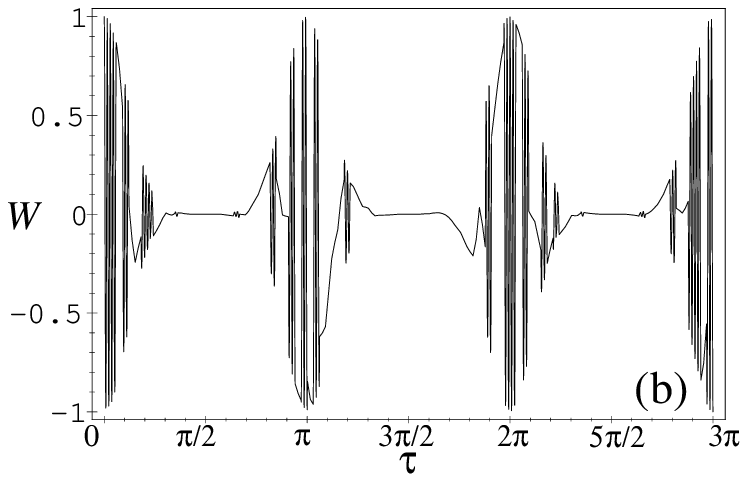}}
\centerline{\epsfxsize=8cm
\epsfbox{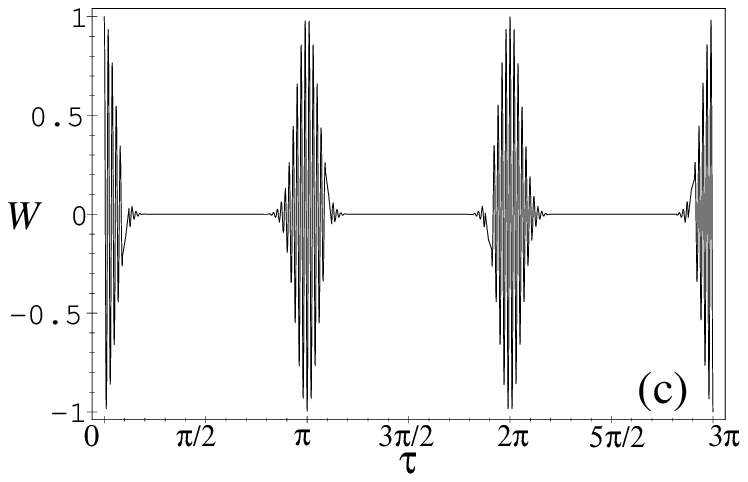}
\epsfxsize=8cm
\epsfbox{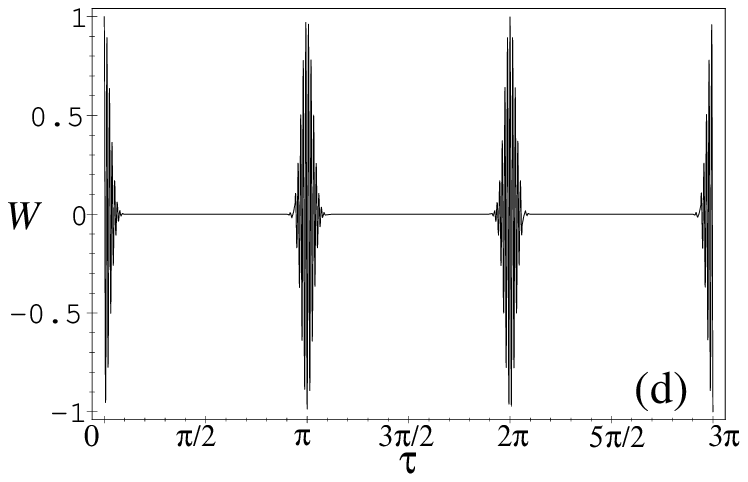}}
\caption{Atomic population inversion as a function of the scaled time
$\tau$. (a) $M=4, \eta=0.999$; (b) $M=70, \eta=0.8$;
(c) $M=70, \eta=0.1$; (d) $M=200, \eta=0.001$.}
\end{figure}

\newpage


\begin{figure}
\centerline{\epsfxsize=8cm
\epsfbox{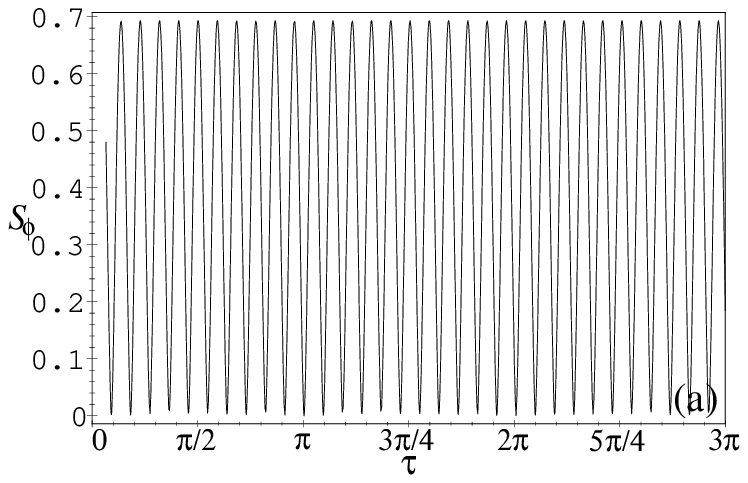}
\epsfxsize=8cm
\epsfbox{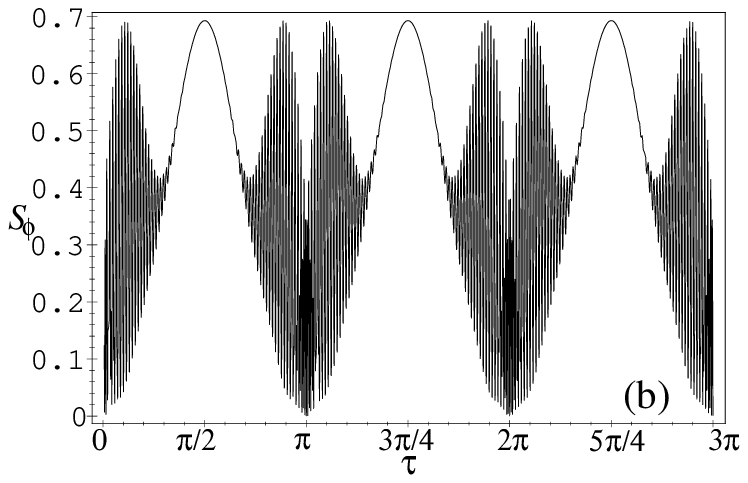}}
\centerline{\epsfxsize=8cm
\epsfbox{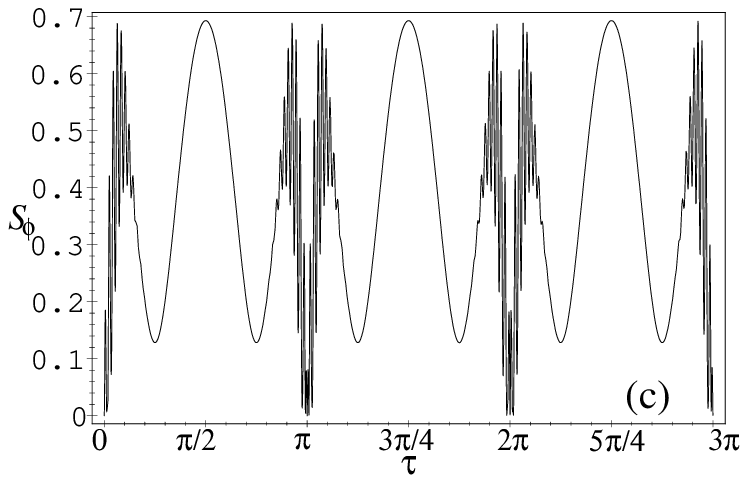}
\epsfxsize=8cm
\epsfbox{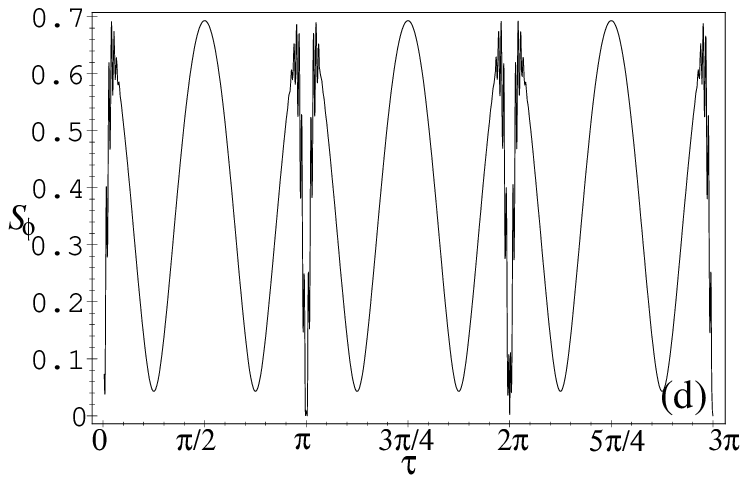}}
\caption{Entropy of the field as a function of scaled time $\tau$. (a) $M=4$
and $\eta=0.9999$ (the initial field state is the number state $|4\rangle$);
(b) $M=70$ and $\eta=0.8$; (c) $M=70$ and $\eta=0.1$; (d) $M=200$
and $\eta=0.005$.}
\end{figure}


\newpage

\begin{figure}
\centerline{\epsfxsize=7cm
\epsfbox{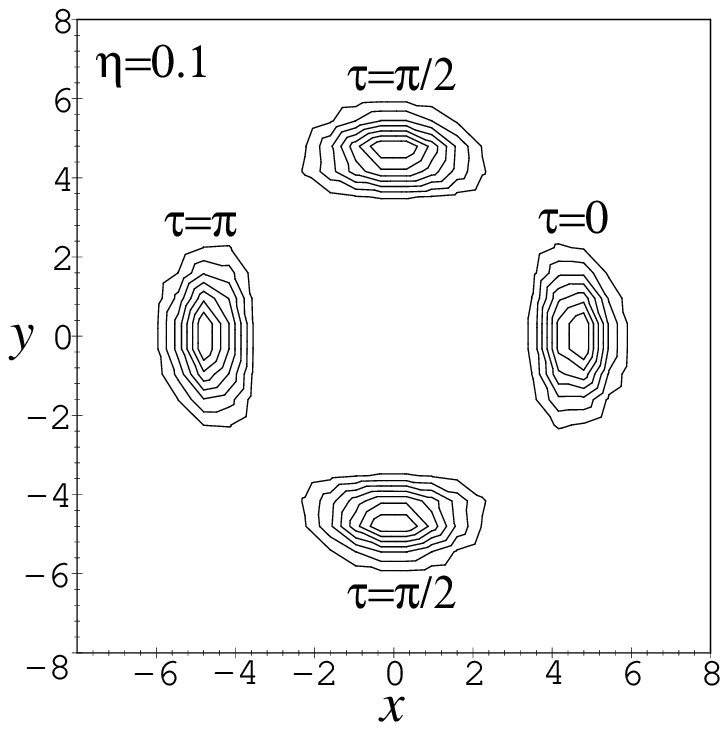}
\epsfxsize=7cm
\epsfbox{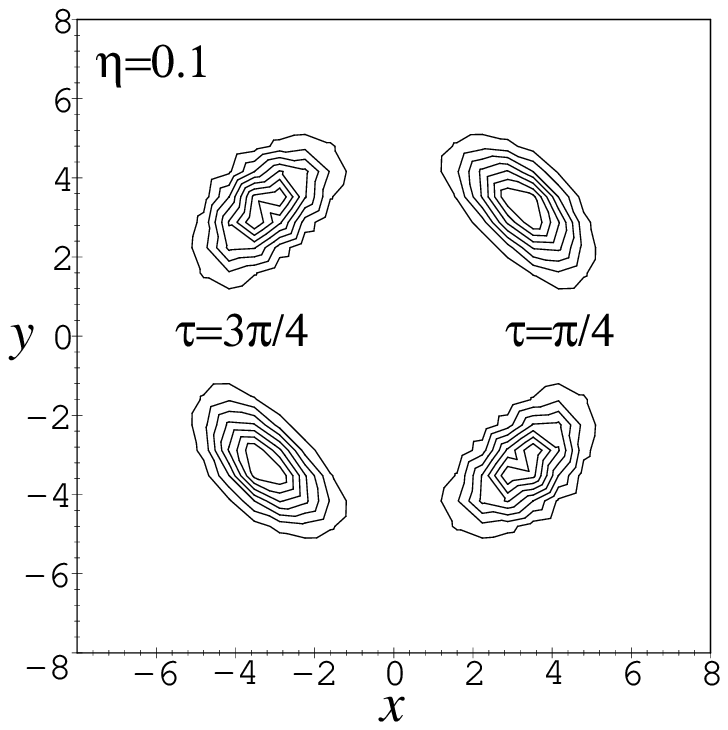}}
\centerline{\epsfxsize=7cm
\epsfbox{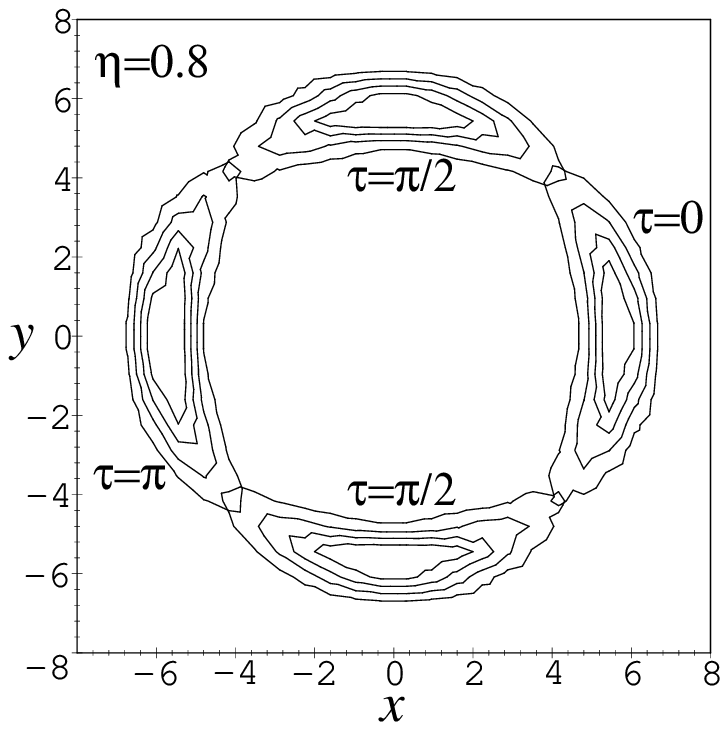}
\epsfxsize=7cm
\epsfbox{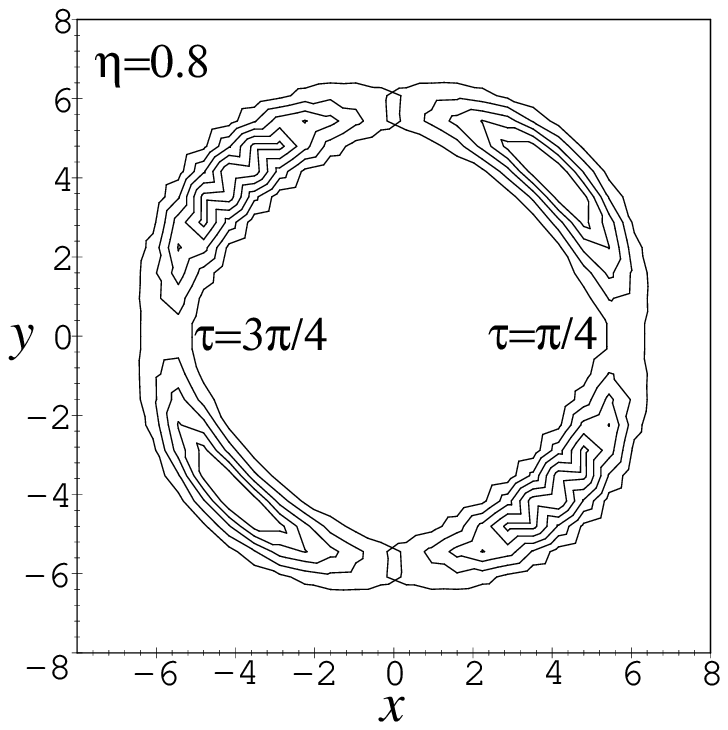}}
\caption{Contour plots of the Q-function of the field at $\tau=0, \pi/4, \pi/2,
3\pi/4$ and $\pi$. Here we choose $\eta=0.1,0.8$.
}
\end{figure}

\newpage

\begin{figure}
\centerline{\epsfxsize=8cm
\epsfbox{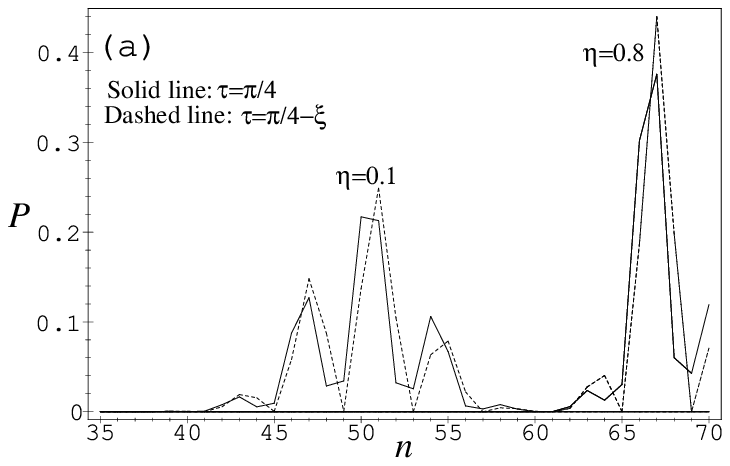}}
\centerline{\epsfxsize=8cm
\epsfbox{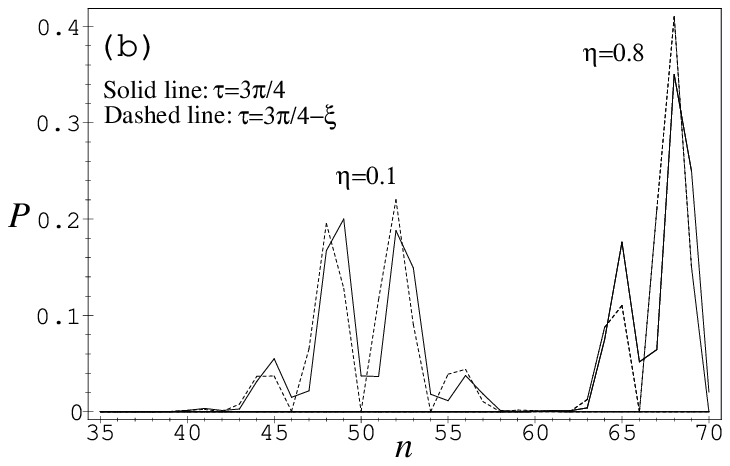}}
\centerline{\epsfxsize=8cm
\epsfbox{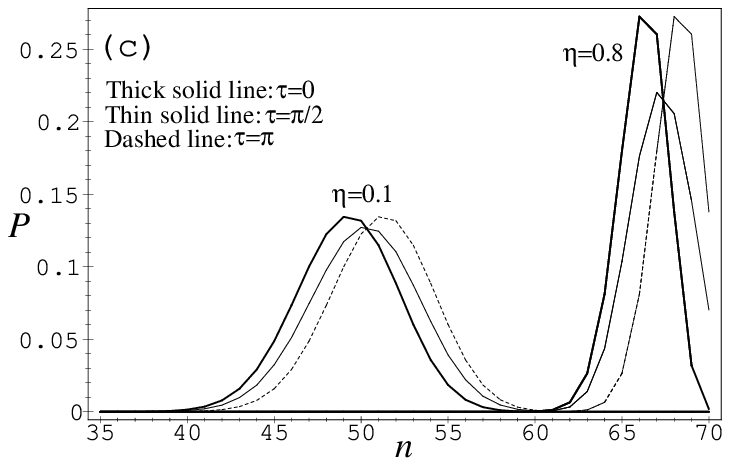}}
\caption{Number distribution of the photon field at different times
$\tau=0, \pi/4, \pi/2, 3\pi/4$ and $\pi$ for $\eta=0.1,0.8$. In
(a) and (b) we also present the distribution at slightly earilier
time $\tau-\xi$, where $\xi$ is chosen as $1/140$ and $1/180$
for $\eta=0.1$ and $\eta=0.8$ respectively.}
\end{figure}

\end{document}